\documentclass{emulateapj}
\usepackage{natbib}
\usepackage{times}
\begin{document}

\shorttitle{The Highest-$z$ GRBs and the SFR}

\shortauthors{Kistler et al.}

\title{The Star Formation Rate in the Reionization Era as Indicated by Gamma-ray Bursts}

\author{Matthew D. Kistler\altaffilmark{1,2},
Hasan Y{\"u}ksel\altaffilmark{3},
John F. Beacom\altaffilmark{1,2,4},
Andrew M. Hopkins\altaffilmark{5}, and
J. Stuart B. Wyithe\altaffilmark{6}}

\altaffiltext{1}{Center for Cosmology and Astro-Particle Physics, The Ohio State University, 191 W.\ Woodruff Ave., Columbus, OH 43210}
\altaffiltext{2}{Dept.\ of Physics, The Ohio State University, 191 W.\ Woodruff Ave., Columbus, OH 43210}
\altaffiltext{3}{Bartol Research Institute and Department of Physics and Astronomy, University of Delaware, Newark, Delaware 19716}
\altaffiltext{4}{Dept.\ of Astronomy, The Ohio State University, 140 W.\ 18th Ave., Columbus, OH 43210}
\altaffiltext{5}{Anglo-Australian Observatory, P.O. Box 296, Epping, NSW 1710, Australia}
\altaffiltext{6}{School of Physics, University of Melbourne, Parkville, Victoria, Australia}


\begin{abstract}
High-redshift gamma-ray bursts (GRBs) offer an extraordinary opportunity to study aspects of the early Universe, including the cosmic star formation rate (SFR).  Motivated by the two recent highest-$z$ GRBs, GRB 080913 at $z \simeq 6.7$ and GRB 090423 at $z \simeq 8.1$, and more than four years of {\it Swift} observations, we first confirm that the GRB rate does not trace the SFR in an unbiased way.  Correcting for this, we find that the implied SFR to beyond $z=8$ is consistent with LBG-based measurements after accounting for unseen galaxies at the faint end of the UV luminosity function.  We show that this provides support for the integrated star formation in the range $6 \la z \la 8$ to have been alone sufficient to reionize the Universe.
\end{abstract}


\section{Introduction}

The connection between gamma-ray bursts\footnote{Throughout, we refer only to ``long'' gamma-ray bursts.} and core-collapse supernovae \citep{Stanek:2003tw,Hjorth} tells us that, in observing a GRB, we are witnessing the death of a massive, short-lived star.  The intense brightness of gamma-ray bursts gives hope that, starting from this principle, we can probe the history of star formation to very early times \citep{Totani,Wijers:1998,Lamb,Porciani,Bromm:2002}, potentially to higher redshifts than with galaxies alone.  First, we must be able to observe the GRBs and obtain redshifts for a sufficient number of events.  Second, we need to understand how to calibrate the GRB rate to the star formation rate (SFR).  {\it Swift}\footnote{See http://swift.gsfc.nasa.gov/docs/swift/archive/grb\_table.} \citep{Gehrels:2004am} has pushed the former greatly ahead, and allowed studies of the latter.

\begin{figure}[b!]
\includegraphics[width=3.25in,clip=true]{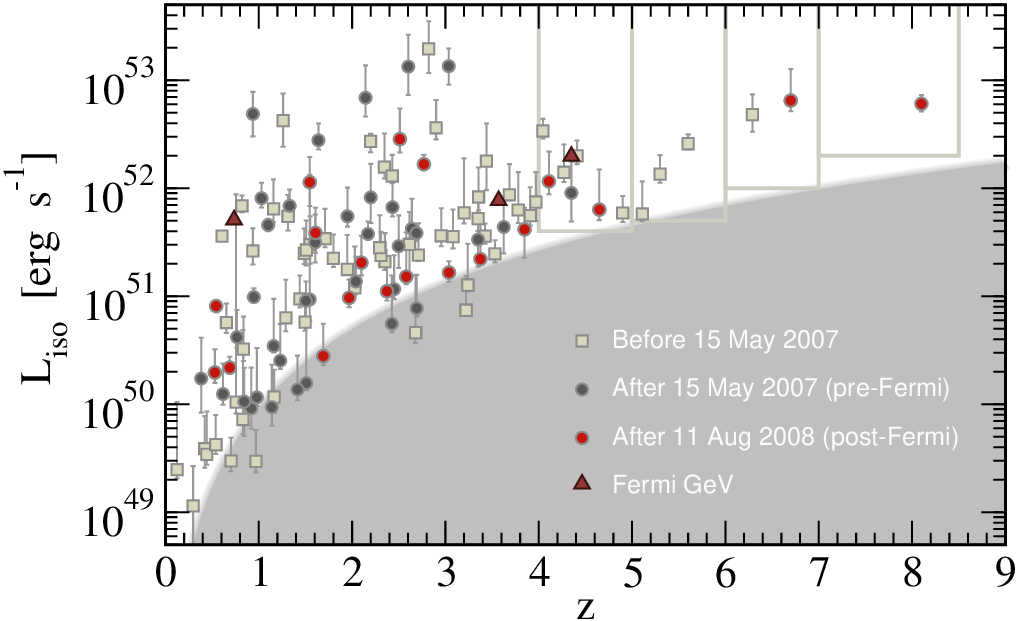}
\caption{The luminosity-redshift distribution of 119 \textit{Swift} GRBs, as we determine from the (updated) \citet{Butler:2007hw} catalog.  Squares represent the 63 GRBs used in \citet{Yuksel:2008cu}, with 56 found subsequently: before (grey circles) and after (red circles) the start of {\it Fermi}.  Three {\it Fermi}-LAT GeV bursts (triangles) are shown (but not used in our analysis).  The shaded region approximates an effective threshold for detection.  Demarcated are the GRB subsamples used to estimate the SFR.
\label{Liso}}
\end{figure}

Our goals are to use the large set of {\it Swift} gamma-ray bursts with known redshifts (see Fig.~\ref{Liso}) accumulated over the last $\gtrsim$ four years to examine the above two points in greater detail.  With improved statistics, we confirm the finding that gamma-ray bursts are not unbiased tracers of the SFR, as in \citet{Kistler} (also see, e.g., \citealt{Daigne, Le:2006pt, Yuksel:2006qb, Salvaterra}), and comment on its suspected origins.  This does not, however, prevent a study of the amount of high-$z$ star formation; it in fact allows for a more proper estimation.

Several recent high-$z$ bursts, most notably GRB 080913 at $z \simeq 6.7$~\citep{Greiner(2009)} and GRB 090423 at $z \simeq 8.1$ \citep{Salvaterra09,Tanvir09}, also allow us to extend the SFR determinations from \citet{Yuksel:2008cu} (which went to $z\sim 6$) to even higher redshifts.  Here, direct SFR measurements are quite challenging, particularly at the faint end of the galaxy luminosity function, where GRBs may be ideal tracers.  Even with only several events, we determine that the SFR declines only slowly from $z\sim 4$ to $z\gtrsim 8$.  This may confirm that a substantial amount of star formation occurs within faint galaxies, in agreement with extrapolations of Lyman Break Galaxy (LBG) measurements, and suggests that stars may be responsible for cosmic reionization.

\section{GRB Sample}

It is easy to understand, with the combination of uncertain extinction corrections, cosmic variance, and selection biases, why measurements of the SFR at high redshifts are difficult endeavors.  Principal among these is that flux-limited surveys observe the bright end of the galaxy luminosity function (LF) and must correct for the faint end, where much of the star formation may be occurring.  The use of gamma-ray bursts as a star formation measure will have its own systematic effects; however the opportunity presented to examine very-high redshifts, and possibly unseen faint galaxies, is great, with no known backgrounds for a bona fide GRB.

To calculate the expected redshift distribution of GRBs, we combine the comoving GRB rate, $\dot{n}_{\rm GRB}(z) = \mathcal{E}(z) \times \dot{\rho}_*(z)$, where $\dot{\rho}_*(z)$ is the SFR density and $\mathcal{E}(z)$ accounts for the fraction of stars resulting in GRBs, with the ability to observe the GRB and obtain a redshift ($0 < F(z) <1$), the fraction of GRBs unobservable due to beaming ($\left\langle f_{\rm beam} \right\rangle$; e.g., \citealt{Racusin(2009),Cenko(2009)}), and the comoving volume per unit redshift\footnote{$dV/dz = 4 \pi \, (c/H_0) \, d_c^2(z) / \sqrt{(1+z)^3\,\Omega_{\rm m} + \Omega_\Lambda}$, where $d_c$ is the comoving distance, $\Omega_{\rm m} = 0.3$, $\Omega_\Lambda = 0.7$, and $H_0 = 70$~km/s/Mpc.} as
\begin{equation}
	\frac{d\dot{N}}{dz} = F(z) 
	\frac{\mathcal{E}(z)\, \dot{\rho}_*(z)}{\left\langle f_{\rm beam}\right\rangle}
	\frac{dV/dz}{1+z}\,.
	\label{dnodz}
\end{equation}
$F(z)$ can be kept constant by considering only bursts with luminosities sufficient to be viewed within an entire redshift range \citep{Kistler}.  We then write $\mathcal{E}(z) = \mathcal{E}_0 (1+z)^{\alpha}$, with $\mathcal{E}_0$ a (unknown) constant that converts $\dot{\rho}_*(z)$ to a GRB rate (in a given GRB luminosity range).  \citet{Kistler} found that $\alpha = 0$ (directly tracing the SFR) was inconsistent with the data at the $\sim 95\%$ level, which favored $\alpha \simeq 1.5$.  As shown in Fig.~\ref{Liso}, many more GRBs have since been detected, warranting a reexamination of this result.

Our sample includes the 63 GRBs used in \citet{Kistler} (up to 15 May 2007), supplemented by 56 subsequent {\it Swift} events with redshifts.  We calculate the intrinsic (averaged) GRB luminosity, $L_{\rm iso} = E_{\rm iso} / [T_{90} / (1+z)]$, from the rest-frame isotropic equivalent (uncorrected for beaming) $1-10^4$~keV energy release ($E_{\rm iso}$) and $T_{90}$, the time interval containing 90\% of the prompt emission, as given in the catalog\footnote{Updated at http://astro.berkeley.edu/$\sim$nat/swift.} of \citet{Butler:2007hw}.\footnote{Note that these values are estimated based upon measurements made by {\it Swift} (in the $15-150$~keV energy band), as explained in \citet{Butler:2007hw}.}  The results for GRBs with $T_{90} > 2$~s are shown in Fig.~\ref{Liso}.

\begin{figure}[t]
\includegraphics[width=3.25in,clip=true]{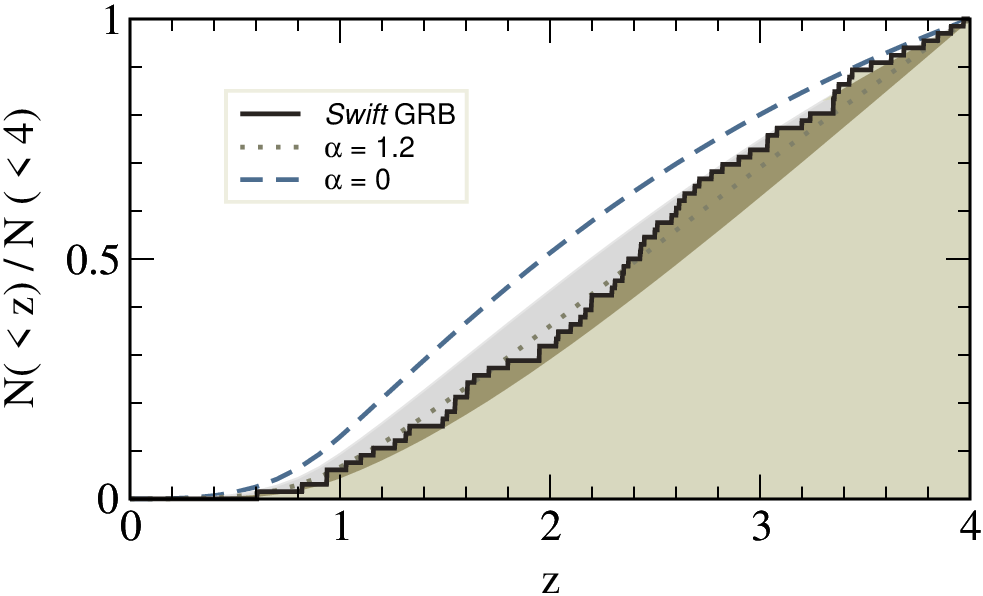}
\caption{The cumulative distribution of the 66 \textit{Swift} GRBs with $L_{\rm iso} >10^{51}$~erg~s$^{-1}$ in $z=0-4$ (solid), as compared to the expectations from the SFH of \citet{Hopkins:2006bw} alone (dashed) and additional evolution of the form $(1+z)^{1.2}$ (dotted).  Outside of the shaded region (bounded by models with $\alpha = 0.6$ and 1.8) corresponds to an exclusion of $> 84\%$.
\label{KS}}
\end{figure}

For this test, we use the cuts defined in \citet{Kistler}: GRBs in the range $z=0-4$ with $L_{\rm iso}>10^{51}$~erg~s$^{-1}$.  This removes many low-$z$, low-$L_{\rm iso}$ bursts that could not have been seen at higher-$z$, leaving us with 66.  The SFR fit from \citet{Hopkins:2006bw} in this range is used as a baseline for comparison, as shown by the dashed line in Fig.~\ref{KS}.  A Kolmogorov-Smirnov test confirms that the GRB rate is incompatible with the expectations from the SFR, now at the $\sim 99\%$ level (possibly higher due to likely missing bursts at $z\lesssim4$) with the present greater statistics, requiring an enhanced evolution relative to the SFR.  Even if we exclude the range $z=1.5-2$, where a larger fraction of redshifts might be missed~\citep{Bloom:2003ic}, the value remains at $\sim 98\%$.  Possible origins of this trend are discussed in detail in \citet{Kistler}, including an overall decrease in cosmic metallicity \citep{Langer}.

Our present result suggests a slightly lower value of $\alpha$.\footnote{This may be due in part to the rate of GRB observations at higher redshifts decreasing noticeably in the period following the cutoff date for our initial GRB sample (for reasons unknown).  Fortunately, as can be seen in Fig.~\ref{Liso}, high-$z$ GRBs detections have since increased (denoted as the period after the start of {\it Fermi} operations).}  Irrespective of the origins of this bias, it must be accounted for in $z=1-4$ to properly relate the GRB rate to the SFR.  In Fig.~\ref{KS}, we show a shaded band bounded above by a model using $\alpha = 0.6$ and below by $\alpha = 1.8$, which can be excluded at $\gtrsim 84\%$.   To be conservative, we will assume this evolution continues to higher $z$, considering $\alpha = 1.2$ throughout.


\section{The High-z Star Formation Rate}

We briefly review the framework laid out in \citet{Yuksel:2008cu} for calibrating a GRB-based estimate of $\dot{\rho}_*$.  This is based on using GRB and SFR measurements in $z = 1-4$ as benchmarks for comparison with bursts of similar luminosity in a higher-$z$ range.  Using only GRBs that could have been detected from anywhere within the volume allows for needed empirical calibration, since neither the conversion from GRB rate to SFR nor the GRB luminosity function are known a priori.  Part of the challenge is in determining the detection threshold versus $z$, since {\it Swift} was designed to maximize GRB detection, though not necessarily in a way well-defined for our purpose \citep{Band:2006fj}.  We show in Fig.~\ref{Liso} an estimated threshold ($\propto d_\ell^2$; see \citealt{Kistler}) based on the GRB luminosities, which acts as a guide to make cuts that maximize statistics and minimize potential ``missing'' bursts.

\begin{figure}[b]
\includegraphics[width=3.25in,clip=true]{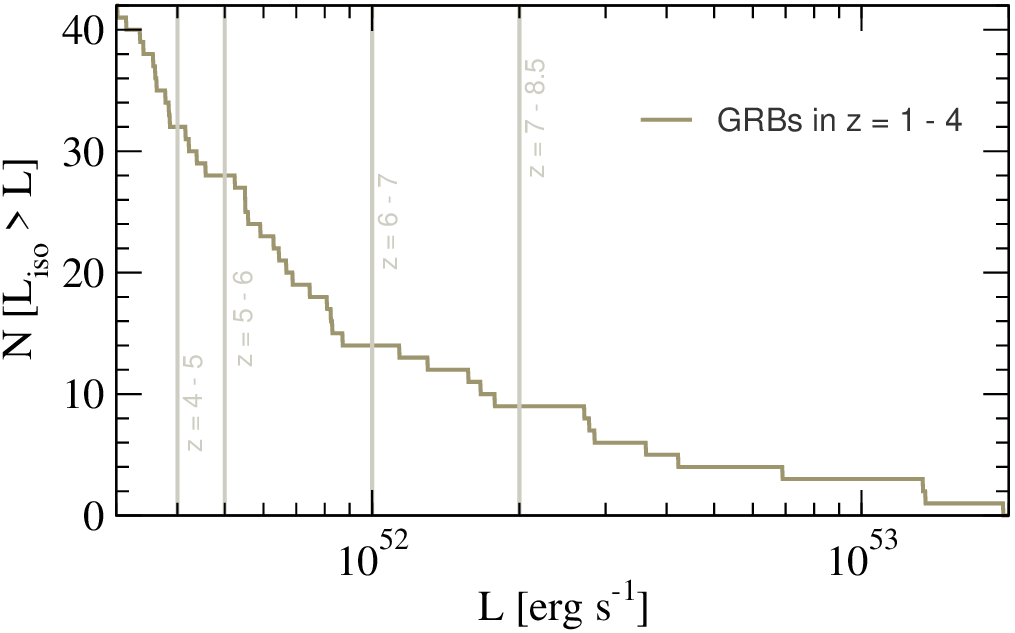}
\caption{The cumulative distribution of $L_{\rm iso}$ for GRBs in the range $z=1-4$.  Shown are the cutoffs used for our GRB subsamples (as in Fig.~\ref{Liso}).
\label{dndL}}
\end{figure}

The cuts and resulting subsamples used for the SFR analysis are shown in Fig.~\ref{Liso}.\footnote{Note that we exclude GRB 060116 (not shown), with a possible photometric redshift of $z=6.6$ (as listed in \citealt{Butler:2007hw}.)}  Fig.~\ref{dndL} shows these in comparison to the distribution of $L_{\rm iso}$ values for bursts in $z=1-4$.  Bursts within each set will be compared to GRBs within the range $z=1-4$ above the given luminosity cut.  We emphasize in advance that, although the final bin contains only GRB 090423 (at $z \simeq 8.1$), even this single event is significant, as it would be quite unlikely if $\dot{\rho}_*$ was too low (see also \citealt{Salvaterra09}).  With this GRB we are entering a regime where the age of the star is becoming non-negligible compared to the age of the Universe, so we extend this bin to $z=8.5$ to cover a plausible range in progenitor lifetime.

The ``expected'' number of GRBs in $z = 1-4$ is
\begin{eqnarray}
\mathcal{N}_{1-4}^{exp}
& = & \Delta t \frac{\Delta \Omega}{4\pi} \int_{1}^{4} dz\,  F(z) \, \mathcal{E}(z)  \frac{\dot{\rho}_*(z)}
{\left\langle f_{\rm beam}\right\rangle} \frac{dV/dz}{1+z} \nonumber \\
& = & \mathcal{A} \, \int_{1}^{4} dz\, \dot{\rho}_*(z)\, (1+z)^{\alpha} \, \frac{dV/dz}{1+z}\,,
\label{N1-4}
\end{eqnarray}
in which $\mathcal{A} = {\Delta t \, \Delta \Omega \, \mathcal{E}_0 \, F_0} / 4\pi {\left\langle f_{\rm beam} \right\rangle}$ depends on the observing time ($\Delta t$), sky coverage ($\Delta \Omega$), and luminosity range of GRBs under examination.  From the average SFR, $\left\langle \dot{\rho}_* \right\rangle_{z_1-z_2}$, the same can be performed for the other ranges as
\begin{eqnarray}
\mathcal{N}_{z_1-z_2}^{exp}
& = &  \left\langle \dot{\rho}_* \right\rangle_{z_1-z_2} 
\mathcal{A} \, \int_{z_1}^{z_2} dz\, (1+z)^{\alpha} \, \frac{dV/dz}{1+z}\,.
\label{Nz1-z2}
\end{eqnarray}
Our interest is in finding $\left\langle \dot{\rho}_* \right\rangle_{z_1-z_2}$ by dividing out $\mathcal{A}$ (using Eq.~\ref{Nz1-z2}).  Taking the measured GRB counts, $\mathcal{N}_{z_1-z_2}^{obs}$, to be representative of the expectations, $\mathcal{N}_{z_1-z_2}^{exp}$, we find
\begin{equation}
\left\langle \dot{\rho}_* \right\rangle_{z_1-z_2} = 
\frac{\mathcal{N}_{z_1-z_2}^{obs}}{\mathcal{N}_{1-4}^{obs}} 
\frac{\int_{1}^{4} dz\, \frac{dV/dz}{1+z} \dot{\rho}_*(z)\, (1+z)^\alpha}{\int_{z_1}^{z_2} dz\,
\frac{dV/dz}{1+z} (1+z)^\alpha}\,.
\label{zratio}
\end{equation}
Note that the decrease of $(dV/dz) / (1+z)$ at $z\gtrsim 1.5$ (as shown in Fig.~1 of \citealt{Kistler}) gives progressively more weight to each observed higher-$z$ GRB.

\begin{figure}[t]
\centering\includegraphics[width=\linewidth,clip=true]{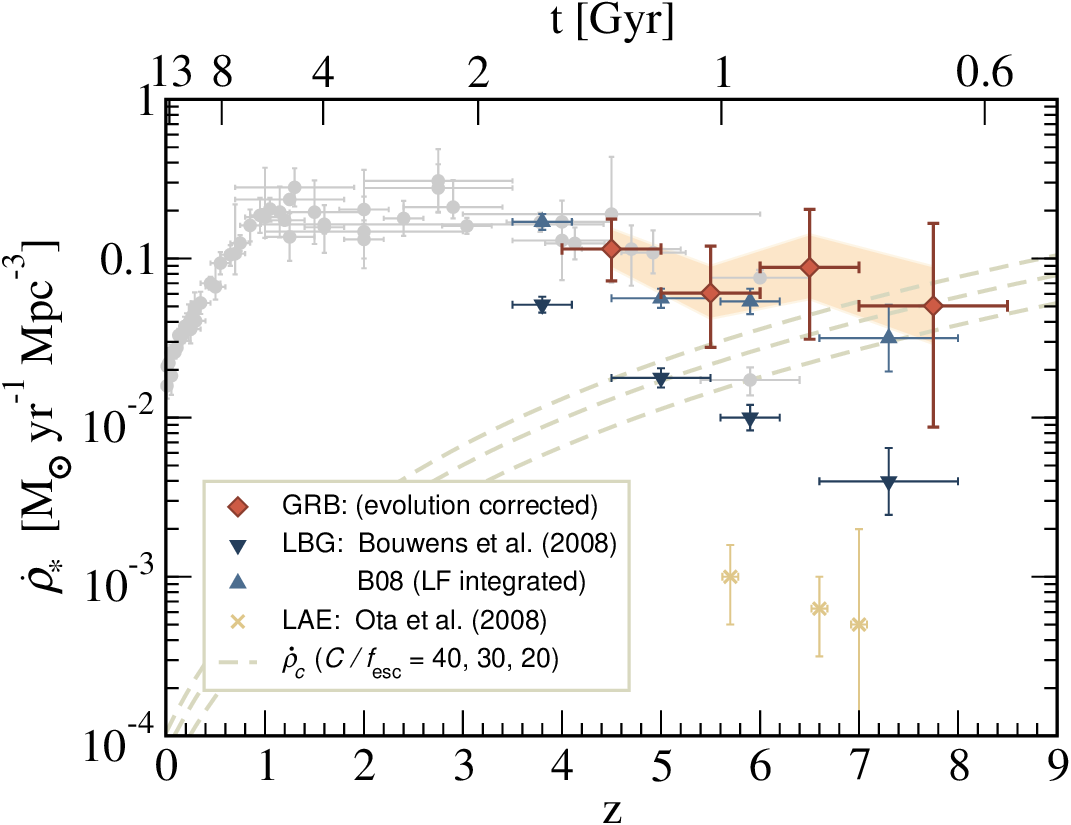}
\caption{The cosmic star formation history.  Shown are the data compiled in \citet{Hopkins:2006bw} (light circles) and contributions from Ly$\alpha$ Emitters (LAE) \citep{Ota et al.(2008)}.  Recent LBG data is shown for two UV LF integrations: down to $0.2 \, L_{z=3}^*$ (down triangles; as given in \citealt{Bouwens08}) and complete (up triangles).  Our (bias-corrected) \textit{Swift} gamma-ray burst inferred rates are diamonds, with the shaded band showing the range of values resulting from varying the evolutionary parameter between $\alpha= 0.6-1.8$.  Also shown is the critical $\dot{\rho}_*$ from \citet{Madau et al.(1999)} for $\mathcal{C}/f_{\rm esc} = 40,\,30,\,20$ (dashed lines, top to bottom).
\label{SFH}}
\end{figure}

We show our new determinations of the high-$z$ SFR in Fig.~\ref{SFH} (assuming a \citealt{Salpeter(1955)} IMF).  Error bars correspond to 68\% Poisson confidence intervals for the binned events \citep{Gehrels:1986mj}.  We also show as a shaded band the values obtained for different assumptions of $\alpha$, bounded above by $\alpha=0.6$ and below by $\alpha=1.8$, which yields an uncertainty smaller than the statistics in the last bins.  Variations due to changing the $L_{\rm iso}$ cutoff can be determined from Fig.~\ref{dndL}, which will typically be less than the statistical uncertainties.  We have been generally conservative and have also verified that using another luminosity estimator, the peak isotropic equivalent luminosity, yields similar results.  Other effects, including the selection of $z$-ranges and the inclusion/exclusion of particular bursts, are discussed in \citet{Yuksel:2008cu}.  We mention only that none of these affect the basic point that the SFR must be large enough to produce the observed GRB counts.

Depending upon the source of the evolution, our bias correction may be unduly underestimating $\dot{\rho}_*$ by a factor of a few at higher $z$.  The most likely astrophysical explanation is due to metallicity.  GRBs are found to favor metal-poor \citep{Stanek:2006gc}, sub-$L_*$ galaxies \citep{Fynbo:2003sx, Le Floc'h:2003yp, Fruchter}, so having a larger fraction of the SFR within such hosts would result in a higher GRB rate.  This could be the case with a steepening faint-end slope of the galaxy LF, so that more of $\dot{\rho}_*$ arises from below $L_z^*$ ($L_*$ as defined at $z$).  This has been observed between $z=0$ and $z\approx 2-3$ (see Fig.~7 of \citealt{Reddy(2009)}).

While our result at $z = 4-5$ is in basic agreement with earlier measurements, at the highest-$z$ ranges, LBG studies probe only the brightest galaxies and must estimate the faint end of the UV LF based on limited data.  Our results diverge from these if corrections for unseen galaxies are not made.  For example, we focus upon the measurements in \citet{Bouwens07,Bouwens08}, which are reported (lower triangles in Fig.~\ref{SFH}) for an integration down to $0.2 \, L_{z=3}^*$ (with their adopted dust corrections).  Fully integrating their UV LFs (which can be regarded as giving a maximum), with faint-end slopes of -1.73, -1.66, -1.74, -1.74 for $\left\langle z \right\rangle=3.8$, 5.0, 5.9, 7.3, respectively, yields the upper set of triangles.

Within the uncertainties, even the highest redshift fully-integrated point now agrees reasonably well with our results, and the preference of GRBs for faint galaxies (although the exact relation between GRB hosts and star forming galaxies as a whole remains to be determined).  We note that the \citet{Bouwens08} LF slope at $\left\langle z \right\rangle=7.3$ was taken to be the same as at $\left\langle z \right\rangle=5.9$.  If the slope is actually steeper (e.g., \citealt{Yan04}), then these measurements could be higher, although it is difficult to draw definite conclusions, due to the limited statistics and uncertainties in dust corrections (e.g., \citealt{Chary05}).


\section{Implications for Reionization}

Transmission in the Gunn-Peterson troughs of high-redshift quasars implies that reionization must have been accomplished before $z = 6$ \citep{Fan et al.(2006)}.  AGN seem to be insufficient for this purpose \citep{Srbinovsky(2007),Hopkins et al.(2008)}, leaving stars as the leading candidate.  To address the ability of an observed population to reionize the Universe, \citet{Madau et al.(1999)} provided an estimate for the required SFR to balance recombination, $\dot{\rho}_c$, which depends upon the fraction of photons that escape their galaxy ($f_{\rm esc}$) and the clumpiness of the IGM ($\mathcal{C}$), updated in \citet{Pawlik(2009)} as
\begin{equation}
	\dot{\rho}_c(z) = \frac{0.027\,M_\odot}{{\rm Mpc}^3\,{\rm yr}}
	\frac{\mathcal{C}/f_{\rm esc}}{30}
	\left[\frac{1+z}{7}\right]^3
	\left[\frac{\Omega_b}{0.0465}\right]^2\,.
	\label{madau}
\end{equation}
For comparison with our empirical SFR, we show in Fig.~\ref{SFH} curves of $\dot{\rho}_c$ as a function of $z$ for $\mathcal{C}/f_{\rm esc}$ = 40, 30, and 20.  We find that our SFR estimates can exceed the $\dot{\rho}_*$ required to keep the Universe ionized at redshifts as high as $z\gtrsim 8$.  However, this criterion refers to an instantaneous equilibrium, and so does not address the requirement that the integrated number of ionizations exceed the number of hydrogen atoms.

\citet{Bolton(2007)} estimated the ionizing emissivity at $z\sim5$ from the Ly$\alpha$ forest, and looked at simple models of the reionization history under different assumptions for the evolution of the ionizing photon emissivity at $z\gtrsim5$.  While their estimate of emissivity from the measured
ionization rate is sensitive to the calculation of mean-free path, \citet{Bolton(2007)} reached the strong conclusion that reionization must have been an extended ``photon starved'' process, and that an emissivity which was constant towards higher $z$ would have been insufficient to reionize the Universe by $z\sim6$.  This implies that the ionizing emissivity must have been higher prior to the end of reionization than just after its conclusion.  The origin of this higher emissivity could lie in an increase in one or all of the SFR, the escape fraction, or the fraction of massive stars in the IMF.   Inspection of Fig.~\ref{SFH} suggests that the SFR is as large at or could even be higher at $z\sim8$ than at $z\sim6$, implying that the ionizing photon emissivity may not be falling towards redshifts greater than $z\sim6$.  Since both ionizing photons and GRBs are produced by massive stars, estimates of the ionizing photon emissivity from the GRB rate should be fairly robust against uncertainties in the IMF at the high-mass end (or at low masses, e.g., \citealt{Wilkins et al.(2008)}).

We are therefore motivated to ask whether we have observed enough star formation at $z\ga6$ to reionize the universe. To answer this question, we make a simple estimate, calculating the number of ionizing photons produced prior to $z\sim6$ given the observed SFR.  For a \citet{Salpeter(1955)} IMF and a metallicity of 1/20 Solar, $\sim 4600$ ionizing photons are produced per baryon incorporated into stars \citep{Barkana:2000fd} (further details are given in \citealt{Wyithe:2009jc}).  Taking this value, together with a constant SFR for a time interval $\Delta t$, we find the number of photons produced per hydrogen in the IGM as
\begin{equation}
  \mathcal{N}_\gamma \sim 4 \left(\frac{f_{\rm esc}}{0.1}\right)
  \left(\frac{\rho_*}{0.1\,M_\odot {\rm Mpc}^{-3}\,{\rm yr}^{-1}}\right)
  \left(\frac{\Delta t}{400\,{\rm Myr}}\right).
\end{equation}
Given our SFR, this implies that $\mathcal{N}_\gamma\sim3^{+3}_{-1.5}(f_{\rm esc}/0.1)$.

In order to reionize the Universe, more than one ionizing photon per baryon is required to compensate for recombinations in the ionized IGM. \citet{Wyithe(2007)} modeled the reionization history including evolution of the clumping factor with the restriction that reionization end at $z\sim6$.  These models yielded $\mathcal{N}_\gamma\sim4$ at $z=6$ under a range of assumptions for the redshift range and efficiency of Population-III star formation.  This is within our estimated range, provided that the escape fraction of ionizing radiation is of order 10\%.  This value of escape fraction lies in the range found by \citet{Srbinovsky(2008)}, who combined semi-analytic models of reionization and the galaxy LF with simulations of the transmission in the high-$z$ Ly-$\alpha$ forest.  While not the final word, our results may thus indicate that stars produced enough ionizing photons in the range $6\la z\la8$ to reionize the Universe.


\section{Discussion \& Conclusions}

With the discovery of the first astrophysical source at $z>8$, \textit{Swift} has enabled GRBs to realize their potential as beacons from the distant past, both into the epoch of reionization and in adequate numbers at lower redshifts to allow for sensible use of the most remarkable events.  Using this wealth of data, we have estimated the star formation rate at the earliest times yet possible, showing that the star formation rate can remain high up to at least $z\sim 8$.  From this, it is plausible that the level of star formation was sufficient to reionize the Universe.

The agreement with direct observations, corrected for galaxies below detection thresholds, suggests that our GRB-based estimates incorporate the bulk of high-$z$ star formation down to the faint end of the LF.  We also see no evidence for a strong peak in the SFR versus $z$.  This assumes that a very strong rise in the efficiency of producing GRBs (beyond that already accounted for), does not hide a drop in the SFR, although this itself would be quite interesting.  While we have not included them in our analysis, of the three {\it Fermi} GeV-detected long GRBs with redshifts (shown in Fig.~\ref{Liso}), two were at $z>3.5$ (e.g., \citealt{Abdo(2009)}).  Their brightness raises the prospect of the independent use of GeV-selected bursts.

The current picture of small, metal-poor GRB hosts observed at low $z$ agrees well with our GRB-inferred SFR being dominated by such sub-$L_*$ galaxies at high $z$.  One might wonder about the whereabouts of these GRB hosts today, whether they continued to grow, merged into more massive halos, etc.  Observations of the afterglow spectrum (e.g., \citealt{Totani(2006),McQuinn:2007gm}) could determine the extent that the host had experienced the effects of the reionizing UV background.  Since GRBs should originate from a different range of overdensities than quasars, potential exists for another examination of the hierarchical history of our Universe.

We thank Kris Stanek for helpful discussions and Nat Butler for making his GRB data available and suggesting peak luminosities.  We acknowledge use of the \textit{Swift} public archive.
MDK and JFB were supported by NSF CAREER Grant PHY-0547102 (to JFB), HY by DOE Grant DE-FG02-91ER40626, and AMH and JSBW by the Australian Research Council.

\clearpage

\end{document}